\documentclass[floatfix,prl]{revtex4}
\usepackage[]{amsmath}

\usepackage[psamsfonts]{amsfonts}
\usepackage[psamsfonts]{amssymb}
	
\usepackage{graphicx}
\usepackage[usenames,dvipsnames]{color}

\newcommand{\comment}[1]{\textcolor{red} {#1}}
\renewcommand{\comment}[1]{\relax}

\newcommand{\Ru} {URu$_{2}$Si$_{2}$}
\newcommand{\threej}[6]{\ensuremath{\left(\begin{array} {ccc}#1 &#2 &#3 \\ #4 &#5 &#6 \end{array}\right)}}

\newcommand{\ninej}[9]{\ensuremath{\left\{\begin{array} {ccc}#1 &#2 &#3 \\ #4 &#5 &#6 \\ #7 &#8 &#9\end{array}\right\} }}

\def\epl{Europhys.\ Lett.}

\def\jpcm{J.\ Phys.: Condens. Matter}
\def\prl{Phys.\ Rev.\ Lett.}
\def\prb{Phys.\ Rev.\ B}

\def\jpsj{J.~Phys.~Soc.~Jpn}

\begin{document}

\title{Systematic study of the hidden order in {\Ru} as a multipolar order; the role of triakontadipoles}

\author{Oscar Gr{\aa}n\"as}
\affiliation{Department of Physics and Astronomy, Uppsala University, Box 516, SE-75120 Uppsala, Sweden}
\author{Francesco Cricchio}
\affiliation{Department of Physics and Astronomy, Uppsala University, Box 516, SE-75120 Uppsala, Sweden}
\author{Lars Nordstr\"om}
\affiliation{Department of Physics and Astronomy, Uppsala University, Box 516, SE-75120 Uppsala, Sweden}


\begin{abstract} A systematic search for possible order parameters of the so-called hidden order of {\Ru} is conducted. Among the possible candidates that do fulfill the experimental symmetry restrictions on the hidden order parameter, we find  one candidate that stand out -- one of the components of a triakontadipole multipole tensor
that belongs to the $A_{1u}$ irreducible representation of the point group $D_{4h}$.  This solution is characterized by a $\vec{Q}=(0,0,1)$ ordering of the triakontadipoles and has a symmetry that forbids magnetic moments as well as most other multipole ordering on the uranium sites. 
This  hidden order phase is closely related to the 
the antiferromagnetic phase, which is manifested in the similarities of the geometries of the calculated Fermi surfaces. 
Finally is is found that this non-magnetic solution allows for another secondary superimposed order parameter that belong to $B_{2u}$, which gives rise to an anisotropic in-plane susceptibility without breaking the tetragonal crystal symmetry.
\end{abstract}

\date{\today}


\maketitle


The  hidden order (HO) of the heavy fermion material {\Ru} has attracted a lot of interest since its discovery in 1985 \cite{Palstra}, as described in a recent review \cite{Mydosh}.
The main enigma is at 17 K there is clear signature of a secondary phase transition but not of any observable order parameter (OP) in the HO phase. The only signal is that there is a tiny staggered magnetic moment which cannot account for the change in entropy at the phase transition. Another important aspect of the HO is that it under pressure goes through a phase transition to an anti-ferromagnetic (AF) phase with relatively large moments. This phase transition is now conclusively established to be of first order \cite{Niklowitz}.
This and the fact that the order parameter of the HO phase does not cause any change in observable symmetries, such as crystal symmetry or magnetic moments, put some severe conditions on its symmetry properties. 

There are numerous theoretical suggestions for the HO parameter in the literature, see Ref.~\cite{Mydosh} for a fuller account of these.  
Some are explicitly itinerant in nature as e.g.~
unconventional density waves \cite{Ikeda}, orbital currents \cite{Chandra:2002p13734}, or helicity order \cite{Varma:2006p365}.
Other focus more on the local atomic order parameter. 
In fact all multipolar order up to rank five have been suggested; quadrupoles \cite{Santini:1994p6460}, octupoles \cite{Kiss:2005p359},
hexadecapoles \cite{Haule-1} and triakontadipoles \cite{Cricchio-URu2Si2}. 

In this Letter we will allow for a general real space atomic order parameter that can arise due to ordering of the itinerant uranium $f$-states. 
First we systematically study what symmetries of the OP that are consistent with various experimental observations.
Then we test which of the OP candidates can be stabilized in a realistic calculations of the correlated $f$-electrons by means of a combined density functional theory and correlation treatment (DFT+$U$). These complementary studies point towards which OP are allowed by symmetry and compatible with the electronic structure of \Ru. 
This study point conclusively towards an OP which is a superposition of two triakontadipole components belonging to different irreducible representations of the isogonal point group: $A_{1u}\oplus B_{2u}$.


We will allow for a staggered OP as a superposition of several independent OP, where each
OP takes a general form
\begin{align} 
 \psi^{\alpha}(\vec{Q}) = \frac{1} {N} \sum_{n}^{N} e^{i\vec{Q}\cdot \vec{R}_{n}} \left< f^{\dagger}_{n}\, \Gamma^{\alpha} f^{\phantom{\dagger}}_{n}\right>
 , \label{genOP}
\end{align}
where  $\vec{Q}$ is an ordering wave vector, $\vec{R}_{n}$ are the uranium atomic positions, $N$ the number of atoms in the crystal,  $f^{\dagger}_{n}$ is the $f$-electron creation operator at atom site $n$ and 
$\Gamma^{\alpha}$ is the operator for the local multipole of type $\alpha$. Here $\Gamma^{\alpha}$ is a matrix-operator in the $14$-dimensional space of $f$-orbitals and $f^{\phantom{\dagger}}_{n}$ is a vector-operator in the same space. 

In order to have  Eq.~(\ref{genOP}) to include all possible OP stemming from the $f$-shell,  $\alpha$ should enumerate all possible degree of freedom within this shell.
This is known to be handled by the so-called  
%
tesseral multipole tensor moments (TMTM)~\cite{Laan-Tmom,Laan:1998,Bultmark-Mult,Cricchio-Pu,Cricchio-URu2Si2,Polarisation}, with $\alpha=\{kpr;t\}$,
\begin{equation}
{w}^{kpr}_{t}(n)=
\mathrm{Tr} \,{\Gamma}^{kpr}_{t} \left< f^{\phantom{\dagger}}_{n}f^{\dagger}_{n} \right>=
 \left< f^{\dagger}_{n}\, \Gamma^{kpr}_{t} f^{\phantom{\dagger}}_{n}\right>\, ,
\label{eq:w_kpr}
\end{equation}
where the corresponding expansion matrices ${\Gamma^{kpr}_t}$  are known matrices in the $f$-orbital space, see also {\em Supplementary Materials} (SM) \cite{SM} for more details. 
The multipole tensors in Eq.~(\ref{eq:w_kpr}) have a simple physical interpretation; for even $k$,  they are multipoles of the charge ($p$=0) 
or spin-magnetization ($p$=1), while for odd $k$ they are multipoles of the corresponding currents. The rank of the tensor is given by $r$ and its time reversal (TR) symmetry is given by $(-)^{k+p}$. 



\comment{ to SM \cite{SM}?:  For a full $f$-shell the TMTM OP with highest rank would be 
of rank 7. 
If the occupation matrix $\left< f^{\phantom{\dagger}}_{n}f^{\dagger}_{n} \right>$ is restricted to the $j=5/2$ states only, the highest rank is 5
and
odd (even) $r$ multipolar moments have to be TR-odd (TR-even). This is not true in the full $f$-shell case.
Additionally, in this case we expect for the spin dependent ($p=1$) TMTM to see a preference for $r=k-1$ tensor moments. They are favored by the spin-orbit coupling for less than half-filled $f$-shell.
Although included in the analysis for completeness, the tensor moments with odd $k+p+r$ are somewhat obscure. Within these the orbital and spin degrees of freedom couple in an axial way, 
as of e.g.~${\mathbf w}^{111}=\frac{2} {3}\vec{\ell}\times\vec{s}$. This tensor moments only arise from the off diagonal $j_{1}\neq j_{2}$ blocks of 
$\left< f^{\phantom{\dagger}}_{n}f^{\dagger}_{n} \right>$. 
They will be included in the analysis and it would be fascinating if they would play a role, but in the present case of {\Ru} they are of marginal interest.}

It is easy to show that  TMTM have simple transformations rules under point group operations. For a rotation with an angle $\phi$ around its quantization axis $c$ and for a rotation by $\pi$ around the first perpendicular direction $a$ they behave as
\begin{align} 
\mathcal{R}(\hat c, \phi)\, {w}^{kpr}_{t}(n)&= \cos(t \phi)\,{w}^{kpr}_{t}(n) + \sin(t \phi)\, {w}^{kpr}_{-t}(n)\nonumber\\
\mathcal{R}(\hat a, \pi)\,{w}^{kpr}_{t}(n)&=(-)^{t+r}\,\mathrm{sgn}(t)\, {w}^{kpr}_{t}(n)\,,
\end{align}
respectively. 
In Table \ref{symm}  the  irreducible representations (IR) of the TMTM are determined for the 
tetragonal crystal point group $D_{4h}$ through the characters of its two generators,
$c_{4}=\mathcal{R}(\hat c, \pi/2)$ and $c_{2}=\mathcal{R}(\hat a, \pi)$, where $a$ and $c$ denote the lattice directions.

\begin{table}
\centering 
\caption{The TMTM components  
 that corresponds to  IR of the group $D_{4h}$. In the enumeration of the components $t$, $n\geq 0$ is an integer.
The experimental compatibility for HO OP belonging to different IR of $D_{4h}$ are ranked from the discussion points {\em i--iv}. Plus (minus) sign means that it has some (dis-) advantageous features for TR-even (g) and TR-odd (u) IR, respectively. For point {\em iii} there is the possibility of superimposed OP and the different letters indicate without any ranking order which two IR have to contribute together.}
\begin{tabular} {|c|r@{}r|c|c|cc|}
\hline  
$D_{4h}$ & cha&racter & even $r$ &{odd $r$} & \quad Hidden & order\quad  \\
 IR & $c_{4}$ & $c_{2}$ &$t$  &$t$ & g & u \\
\hline
$A_{1}$ &  1& 1 & $+4n$ &  $-4(n+1)$ &++a$-$ & ++b+ \\
$A_{2}$ &  1& $-1$  & $-4(n+1)$  & $+4n$ &++c+ & +$-$d$-$ \\
$B_{1}$ &  $-1$& 1  & $+2(2n+1)$  & $-2(2n+1)$ & ++c$-$ & ++d$-$ \\
$B_{2}$ &  $-1$&$ -1$  & $-2(2n+1)$ & $+2(2n+1)$ & ++a$-$ & ++b$-$   \\
$E$ & 0 & 0 & $\pm (2n+1)$  & $\pm (2n+1)$  &$-$++$-$ &$-$$-$+$-$ \\
\hline 
\end{tabular}
\label{symm}
\end{table}

Before we present our calculations we will reinvestigate which of the TMTM OP that are compatible with the most important constraints put on the HO OP that have been gathered by the huge amount of experimental studies. It is not possible to cover all experimental aspects \cite{Mydosh}, so we concentrate on experimental observations that have simple and direct implication on the symmetry aspects of the HO. 

{{i}) {\bf Non-broken lattice symmetry.}} At high temperatures {\Ru} has the crystal symmetry of the space group 139 (I4/mmm) with its isogonal point group $D_{4h}$ (4/mmm). It is established that there are no lattice distortions at the HO transition at 17 K, i.e.~the HO phase also belong to the tetragonal crystal class. Since the uranium atoms are situated at maximally symmetric Wyckoff sites the local site symmetry is also $D_{4h}$. Then from Table \ref{symm} it is clear that if the local four-fold rotation remains, the OP has to belong to the $A_{1}$ or $A_{2}$ IR of the point group. However when the crystal symmetries are taken into account one can see that the broken local four-fold symmetry can be taken care of by crystal symmetry operations, which has been noted earlier for the case of quadrupoles \cite{Harima:2010}.
The remedy is  a four-fold screw axis generated by the non-symmorphic symmetry operation $(c_{4}|\frac{1} {2}\frac{1} {2}\frac{1} {2})$.
In the cases of $B_{1}$ and $B_{2}$ IR the tetragonal symmetry is recovered for an ordering wave vector $\vec{Q}=(001)$, while for $E$  the corresponding wave vector is $(00\frac{1} {2})$.


{{ii})  {\bf Vanishing magnetic moments.}}
Since it is now established that the phase transition under pressure from the HO phase to the AF phase is of first order \cite{Niklowitz}, the HO OP cannot belong to $A_{2u}$ since it is the IR of magnetic moments along the $c$-axis. 
If we are looking for a symmetry reason for the non-existence of magnetic moments in the HO phase, the IR of the HO OP 
cannot neither belong to $E_{u}$, the IR of the in-plane magnetic moments. 

{iii})  {\bf Low symmetry in-plane susceptibility.}
Recently there have been reported that the HO phase has a broken four-fold symmetry \cite{Okazaki}.
An analysis has shown that this arises from the  OP squared that interacts with the applied magnetic field squared, 
which leads to a non-vanishing $B_{2g}$ IR of the magnetic susceptibility \cite{Thalmeier,SM}.
In Ref.~\cite{Thalmeier} it was shown that only the OP belonging to $E$ has a $B_{2g}$ IR in the direct product representation of its square, $B_{2g}\in E_{\nu}\otimes E_{\nu}$, $\nu=\{g,u\}$. 

Another option is that OP is a linear combination of two independent OP of different IR. Then since \cite{SM}
$A_{1\nu}\otimes B_{2\nu}=A_{2\nu}\otimes B_{1\nu}= B_{2g}$, we see that the co-existence of these type of OP would also lead to the observed variation of the magnetic susceptibility. Note that the two OP have to have the same TR symmetry $\nu$, i.e.~even or odd.

{{iv})  {\bf Good hideout.}}
From Table \ref{symm} we can observe that even when the primary OP would have a high rank which makes it hard to observe directly, 
it will leave traces  in terms of induced multipoles of lower ranks. In that sense there
are two type of OP that have the chance to be better hidden than others.
They are OP with $t=-4$  and either even $r\geq 4$ belonging to IR $A_{2}$ or with odd $r\geq 5$ belonging to IR $A_{1}$.

 {\bf Summary of candidates.}  Our survey is summarized in Table \ref{symm}. It is a linear combination of OP that best fulfill the criteria from point {\em i}-{\em iv}, either
$A_{1u}\oplus B_{2u}$ or $A_{2g}\oplus B_{1g}$, while
the best pure OP  belongs to either $A_{1u}$ or $A_{2g}$. 

{\bf Electronic structure calculations.}
In order to determine which of the TMTM are compatible with the electronic structure, we will perform a systematic survey in terms of realistic calculations.
Care has to be taken to allow for TMTM OP solutions as of Eq.~(\ref{genOP}) that belong to the different IR of the isogonal group $D_{4h}$ as listed in Table \ref{symm}. In our approach we break the symmetry by inducing staggered  $w^{kpr}_{t}$ on the uranium sites and determine the largest possible symmetry group that is compatible with their existence. Then by iteration we determine if this starting assumption converges to a non-trivial solution. 
In principle it is possible that two cases of induced OP of the same IR do lead to two different solutions, but they may of course also converge to the same solution.

The electronic structure is determined with the DFT+$U$ approximation within the APW+$lo$ method as implemented in the {\sc elk}-code \cite{LAPW,Elk}.
The calculations were performed in a similar way as earlier described  \cite{Bultmark-Mult,Cricchio-URu2Si2,Polarisation}, and is presented in more details in SM \cite{SM}. 
\comment{to SM \cite{SM}?: In the DFT+$U$ approach a screened Hartree-Fock interaction is included among the 5$f$ states only. The Slater parameters $F^{(k)}$ are determined individually by using a screened Yukawa potential, this approach has been showed to be particularly convenient since all four $F^{(k)}$ are controlled by the choice of a single parameter, 
the parameter $U$ \cite{Bultmark-Mult}.
For double-counting we adopt the automatic interpolation scheme of Pethukov {\em et al} \cite{Pethukov-INT}.
The calculations were performed at the experimental lattice constants \cite{Palstra}. 
The muffin-tin radius $R^{U}_{\mathrm{MT}}$ of U is set to 1.9~\AA, and those of
Ru and Si are set to 1.2 {\AA}. The parameter $R^{U}_{\mathrm{MT}} |\vec{G}+\vec{k}|_{\mathrm{max}}$, governing  the number of plane waves in the APW+$lo$ method, is chosen to be 9.5. To allow for the $\vec{Q}$ order the unit cell is doubled and the corresponding Brillouin zone is sampled with $18\times18\times10$ $k$-points.}

For each possible TMTM component of each IR  as given in Table \ref{symm} we have started a calculation with a large value of the corresponding multipole. 
All calculations enforce a staggering wave vector $\vec{Q}=(001)$ for tensors with rank $r\ge1$.
 It was found that one solution exists for each TR-labelled IR and these results are collected in Table \ref{calc} for the case of $U=1$ eV. 
In order to quantify the importance of different tensor components we have utilized the concept of polarization $\pi^{kpr}_{t}$ \cite{Polarisation}, which
 is a normalization independent quantity that directly measures the importance of the different contributions to the polarization of the density matrix. It is proportional to the square of the components of the TMTM $w^{kpr}_{t}$ and all components except $kpr=000$
add up to a total polarization $\pi^\mathrm{tot}$, see SM \cite{SM} for more details. 

In the case of TR-even all IR converged to the trivial un-polarized case of $A_{1g}$ with only the rotational invariant tensors, $w^{000}_{0}$ and $w^{110}_{0}$, being non-zero. They correspond to expectation values of  the $f$-occupation and the operator $\ell\cdot s$, respectively.  The calculated large value of $w^{110}_{0}$ leads to an enhancement of the intrinsic spin-orbit coupling, which brings the solution in to a relativistic regime with predominantly $j=5/2$ occupation. The resulting OP is not actually staggered. 

When the TR symmetry is broken we find in all cases that components of the triakontadipole $w^{615}_{t}$ dominate. In addition we find smaller contributions from all TMTM components that are allowed by symmetry as given in Table \ref{symm}, that is both TR-odd as well as TR-even. However, only two TMTM have significant polarizations,
$w^{110}_{0}$ and $w^{615}_{t}$ for all cases except $A_{2u}$ and $E_{u}$. 
In all these cases the $w^{110}_{0}$ have similar polarizations as in the TR-even case.

The case of $A_{2u}$ is the most complicated. Here there is a strong competition for the exchange energy between the magnetic dipole moments along $c$-axis and the two allowed triakontadipole components.
The magnetic dipoles have significant polarization for all three MP variants, spin $w^{011}_{0}$, orbital $w^{101}_{0}$, and spin dipolar $w^{211}_{0}$, while the triakontadipole $w^{615}_{0}$ has a contribution  of similar strength but 
$w^{615}_{4}$ has the largest polarization. Although all initializations lead to the same solution, the convergence is often extremely slow which indicates that there is a flat energy-landscape.
This was observed in the earlier study where we first observed the triakontadipoles in the $A_{2u}$ phase \cite{Cricchio-URu2Si2}.
%
For the two dimensional IR $E_{u}$ we have performed two calculations where the OP is oriented along the in-plane symmetry directions (100) and (110), respectively.
Both the two solutions are stable although different, indicating a strong anisotropy. As anticipated both cases posses non-vanishing magnetic  moments. The local magnetic moments are 1.1 and 0.8 $\mu_\mathrm{B}$, respectively. 
No polarized solution could be found in the case of tetragonal $E_{u}$, i.e.~with the ordering vector $\vec{Q}/2$.

For the TR-odd case we have also looked for the possibility of superimposed OP.  
Only one new solution was found  when allowing for various combinations of superimposed OP, 
but this is one of the HO OP candidates that were singled out fulfilling all the experimental constrains.
It is $A_{1u}\oplus B_{2u}$, with the $\psi^{615}_{-4}(\vec{Q})$ two orders of magnitude larger than $\psi^{615}_{2}(\vec{Q})$.
The product of these two OP interacts with the global magnetic field in the $ab$-plane and gives rise to a nonzero $B_{2g}$ IR of the magnetic susceptibility \cite{SM}, which explains the in-plane response of the torque experiments \cite{Okazaki}. 


\begin{table}
 \centering 
 \caption{The different solutions 
for calculations with $U=1$ eV.  For each IR label, the total polarization, the dominating ($\pi^\mathrm{kpr}_{t}\geq 1$) TMTM components, their values and their polarizations are given. For $E_{u}$ the case with largest polarization, which is with the two independent components along the plane diagonal (110), is presented. \label{calc}}
  \begin{tabular} {|c|c|c@{ }c|c|c|}
\hline  
 IR    &  $\pi^\mathrm{tot}$ &$kpr$ & $t$ & $w^\mathrm{kpr}_{t}$& $\pi^\mathrm{kpr}_{t}$ \\
\hline
$A_{1g}$   & 4.8  & 110 & 0 & $-2.5$ & 4.8 \\
\hline
 $A_{1u}$  & 12.8 & 615 & $-4$ & 44.9 & 7.0  \\
\hline
$A_{2u}$  & 14.6  & 011 & 0& 1.0 & 1.0\\
 & & 101 & 0 & $-0.8$ & 1.3 \\
 & & 615 & 4& $-36.5$ & 4.7\\
\hline
$B_{1u}$  & 12.2  & 615 & $-2$ & 39.7 & 5.5 \\
\hline
$B_{2u}$ & 11.9  & 615 & $2$ & 39.6 & 5.5 \\
\hline
$E_{u}$ (110) & 14.5 & 615 & $\pm 5$ & $-24.2$ &  2.0 \\
   & & 615 & $\pm 1$ &   20.3 & 1.4 \\
\hline
\hline
$A_{1u}\oplus B_{2u}$  & 12.8 &615 & $-4$ & 44.9 & 7.0  \\
  &  &615 & $2$& 0.5 & $7\cdot 10^{-4}$\\
\hline 
\end{tabular}
\label{calcOP}
\end{table}


\begin{figure}[htbp]
\begin{center}
\includegraphics[width=1.0\columnwidth]{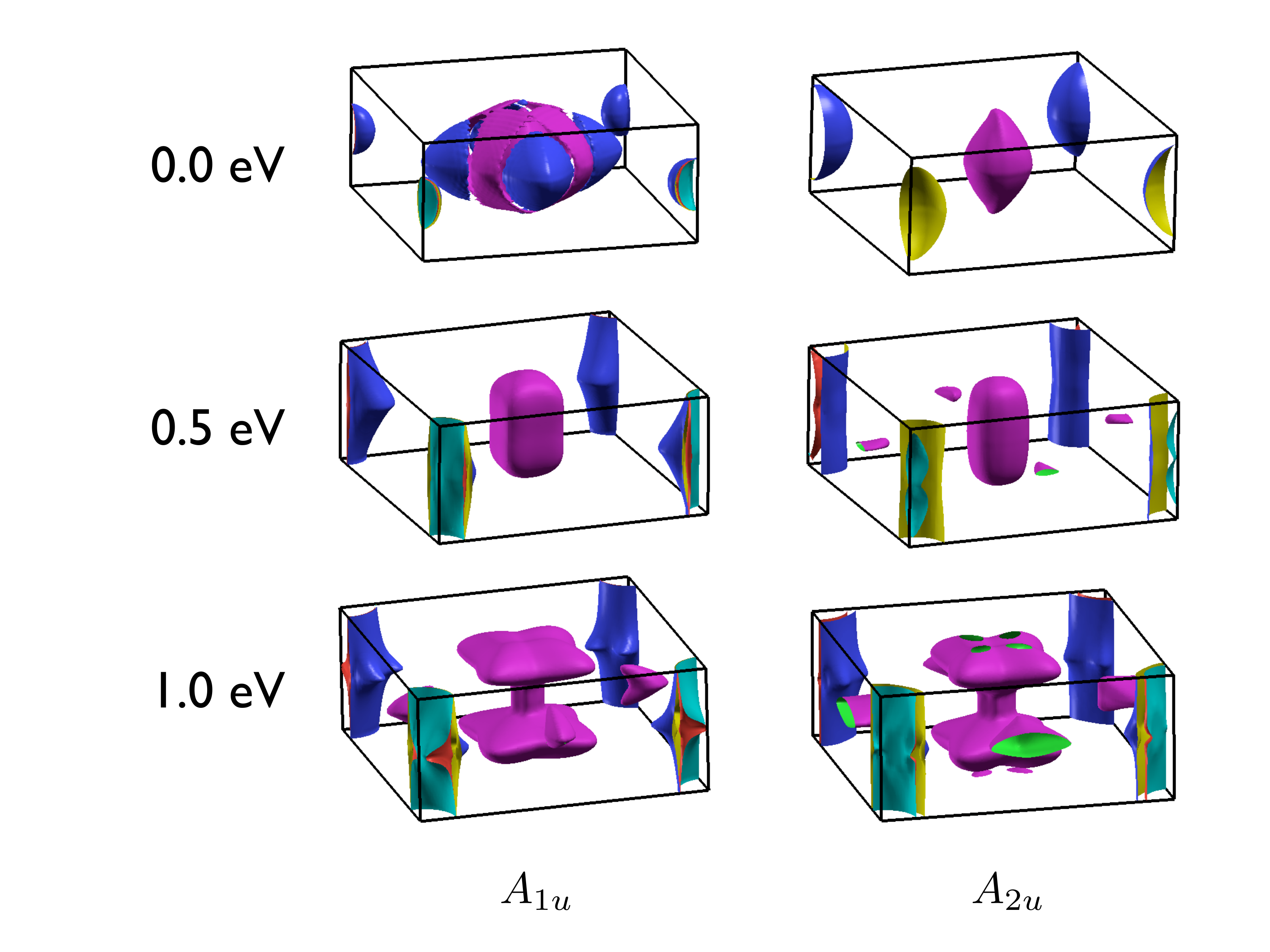}
\caption{The Fermi surface of  {\Ru}  as a function of parameter $U$ for the case of OP belonging to the IR $A_{1u}$ and $A_{2u}$, respectively.
The green/purple, blue/yellow and red/light-blue colors on the different sides of the Fermi sheets refer to the electron/hole side. 
      \label{FS}}
\end{center}
\end{figure}

In Fig.~\ref{FS}   the variation of the Fermi surface (FS) sheets  are displayed, from the uncorrelated cases to the finite $U=0.5$ eV and $1.0$ eV cases for both the two solutions with largest polarizations, the $A_{1u}$  and $A_{2u}$. At $U=0$ they reproduce the FS of Ref.~\cite{Elgazzar-URu2Si2}. $A_{2u}$ corresponds to their AF solution, and the, in this case unpolarized, $A_{1u}$ corresponds to their paramagnetic. In this limit, we observe that the FS of the two solutions are radically different, while at finite $U$ they become surprisingly similar. 
Hence if we would identify the non-magnetic solution with a $A_{1u}$ OP as the HO phase and the magnetic solution $A_{2u}$ as the AF phase, we see that the calculated FS are 
 in excellent qualitative accordance with recent Shubnikov-de Haas measurements~\cite{Hassinger-URu2Si2} that observe only minor changes in the phase transition from HO to AF phase under pressure.
 In the calculated case we can see very  large resemblances in topology as well as sizes between the FS of the two solutions at both $U=0.5$ eV and $U=1$ eV. Since the FS for the two different  $U$-values do not even have the same topology, the FS geometry is very $U$ dependent.  So it is left for a future study to calculate more in detail the $U$ variation of the FS and to compare more quantitatively with experiments.

In this study we have performed a detailed and systematic survey of possible HO OP in terms of ordering within the uranium $f$-bands. 
A  picture arises from some common features of Tables \ref{symm} and \ref{calc}. First we observe there are no indication at all in the calculations for a TR-even HO OP.
Secondly, it is clear that the AF solution at high pressure is the IR $A_{2u}$ where the OP in the calculation is a superposition of magnetic vector OP 
and TR-odd rank 5 TMTM components, 
with the last ones dominating.
Thirdly, this phase competes with an almost pure TMTM OP of the kind $\psi^{615}_{-4}(\vec{Q})$ which belong to the IR $A_{1u}$.
 These two solutions, $A_{2u}$ and $A_{1u}$, are among the ones that have the largest polarizations, as shown in Table \ref{calc}. 
The HO phase $A_{1u}$ only allows a few multipolar components with tesseral component $t=-4$, where all except $w^{615}_{-4}$ become small. This leads to an optimally  hidden OP. It is further found that there exist a solution where this $A_{1u}$ OP is superimposed with a smaller $B_{2u}$ triakontadipole OP, which would explain the recently
observed
anisotropic in-plane susceptibility\cite{Okazaki}.


From Table \ref{calcOP} we can directly observe that the various solutions for the different IR  all have a large contribution from the rank 5 TMTM OP $\psi^{615}$. 
These results are in good accordance with our earlier observations that these trikontadipoles play a large role for {\Ru}  in particular  \cite{Cricchio-URu2Si2} and for the time reversal symmetry breaking for moderate or strong spin-orbit coupling in general  \cite{Polarisation}. The latter is summarized as Katts' rules. 
After completion of this study we became aware of a recent calculational study \cite{Ikeda-12} which also identify the importance of large rank 5 multipoles in the electronic structure of \Ru. In those random phase approximation (RPA) calculations only the $j=5/2$ states  are included (see \cite{SM} for a discussion) and it was concluded that the magnetic $r=5$ $E_{u}$ state is the best candidate for the HO phase.

The support  from 
the Swedish Research Council (VR) is thankfully acknowledged. 
The calculations have 
been performed at the Swedish high performance centers 
HPC2N and NSC 
under grants provided by the 
Swedish National Infrastructure for Computing (SNIC).

\newpage

\begin{center}{\LARGE\sc Supplementary Materials}\end{center}

\section{Order parameters}
In this study our main focus is on order parameters (OP) of the general form
\begin{align} 
 \psi^{kpr}_{t}(\vec{Q}) &= \frac{1} {N} \sum_{n}^{N} e^{i\vec{Q}\cdot \vec{R}_{n}} \left< f^{\dagger}_{n} \Gamma^{kpr}_{t} f^{\phantom{\dagger}}_{n}\right>\nonumber\\
 &=\frac{1} {N} \sum_{n}^{N} e^{i\vec{Q}\cdot \vec{R}_{n}}\, w^{kpr}_{t}(n)
 , \label{genOP}
\end{align}
where  $\vec{Q}$ is an ordering wave vector, $\vec{R}_{n}$ are the uranium atomic positions, $N$ the number of atoms in the crystal and $f^{\dagger}_{n}$ is the $f$-electron creation operator at atom site $n$.
$\Gamma^{kpr}_{t}$ is the operator for the local tesseral multipole tensor moment (TMTM) component, 
\begin{equation}
{w}^{kpr}_{t}(n)=\mathrm{Tr} \, {\Gamma}^{kpr}_{t}\rho_{n} =
\mathrm{Tr} \,{\Gamma}^{kpr}_{t} \left< f^{\phantom{\dagger}}_{n}f^{\dagger}_{n} \right>=
 \left< f^{\dagger}_{n}\, \Gamma^{kpr}_{t} f^{\phantom{\dagger}}_{n}\right>\ ,
\label{eq:w_kpr}
\end{equation}
where the trace is over the $f$ orbitals.
For a $f$-shell $0\le k\le 6$, $0\le p \le 1$ and $|k-p|\le r\le k+p$, which constitute 26 different multipole tensors, of which 13 are time reversal (TR) even and 13 TR-odd,
and the total number of tensor components are $196=14\times 14$. This then accounts for the full freedom of the 14-dimensional density matrix $\rho_{n}=\left<f^{\phantom{\dagger}}_{n}f^{\dagger}_{n} \right>$. 
In the  $14$-dimensional space of $f$-orbitals, $\Gamma^{kpr}_{t}$ is a matrix-operator  and $f^{\phantom{\dagger}}_{n}$ is a vector-operator. 
{In a $\{jm_{j}\}$-representation of the $f$-states ($jj$-basis) we have
\begin{align}
{\Gamma}^{kpr}_{t,12}=&
\frac{\sqrt{[j_{1}j_{2}]}} {N_{kpr\ell}}(-)^{k+p+r}
 \ninej{\ell} {\ell} {k} {s} {s} {p} {j_{1}} {j_{2}} {r} \, \gamma_{t,m_{1}m_{2}}^{j_{1}j_{2}r}\label{Gamma}\\
 \gamma_{t,m_{1}m_{2}}^{j_{1}j_{2}r}=&(-)^{j_{1}-m_{1}}\mathcal{T}  \threej{j_{1}} {r} {j_{2}} {-m_{1}} {t} {m_{2}}\,.
\label{gamma}
\end{align}
Here 
$\ell=3$, $s=1/2$, 
the $\left(\dots\right)$- and $\left\{\dots\right\}$-symbols are the Wigner-3j and -9j, respectively, 
$N_{kpr\ell}$ is a normalization factor
and $[a...b]=(2a+1)...(2b+1)$. \cite{Bultmark-Mult,Laan-Tmom}
The operator $\mathcal{T}$ brings a spherical tensor, which was used in earlier studies \cite{Cricchio-Pu,Bultmark-Mult}, to a tesseral form}
\begin{align} 
\mathcal{T} a_{t}=
\left\{\begin{array} {lr} 
\left[{a}_{t}+(-)^t
  {a}_{-t}\right]/\sqrt{2}=\sqrt{2}\, \Re\, a_{t}& t> 0\\
{a}_{t}& t=0\\
  i\left[{a}_{t}-(-)^t
  {a}_{-t}\right]/\sqrt{2}=\sqrt{2}\, \Im\, a_{|t|}& t< 0\\
  \end{array}\right..
\end{align}
The tesseral form is convenient when considering rotational symmetries as in the present study.
The TMTM in Eq.~(\ref{genOP}) have a simple physical interpretation; for even $k$,  they are multipoles of the charge ($p$=0) 
or spin-magnetization ($p$=1), while for odd $k$ they are multipoles of the corresponding currents. The rank of the tensor is given by $r$ and its time reversal (TR) symmetry is given by $(-)^{k+p}$.

Hence all possible OP stemming from the $f$-shell is covered by a superposition of OP in terms of TMTM of Eq.~\ref{genOP}.
For instance the OP of an ordinary spin density wave
is given by $\psi^{011}_{t}(\vec{Q})$.  This can be easily seen since $\Gamma^{011}_{t}=\tilde{\sigma}_{t}$, the Pauli spin matrices in tesseral form, i.e.~$\tilde{\sigma}_{1}=\sigma_{x}$, $\tilde{\sigma}_{-1}=\sigma_{y}$ and $\tilde{\sigma}_{0}=\sigma_{z}$, respectively.
Another example is $\psi^{112}_{2}(\vec{Q})$, which is one of the staggered quadrupoles models suggested in Ref.~\onlinecite{Santini:1994p6460}.

\subsection{Symmetry properties}
It is easy to show that  TMTM have simple transformations rules under point group operations. For a rotation with an angle $\phi$ around its quantization axis (which we denote $z$) and for a rotation by $\pi$ around the first perpendicular direction (which we denote $x$) they behave as
\begin{align} 
\mathcal{R}(\hat z, \phi)\, {w}^{kpr}_{t}(n)&= \cos(t \phi)\,{w}^{kpr}_{t}(n) + \sin(t \phi)\, {w}^{kpr}_{-t}(n)\nonumber\\
\mathcal{R}(\hat x, \pi)\,{w}^{kpr}_{t}(n)&=(-)^{t+r}\,\mathrm{sgn}(t)\, {w}^{kpr}_{t}(n)\,,
\end{align}
respectively. Hence the rotational properties are determined by the rank $r$ and the component $t$ only. In Table I of the main Letter (ML) the  irreducible representations (IR) of the TMTM are determined for the 
isogonal point group $D_{4h}$ through the characters of its two generators,
$c_{4}=\mathcal{R}(\hat c, \pi/2)$ and $c_{2}=\mathcal{R}(\hat a, \pi)$.
 In addition the TMTM behave under a TR-operation $\Theta$ as
\begin{align} 
\Theta\,{w}^{kpr}_{t}(n)&=(-)^{k+r}\, {w}^{kpr}_{t}(n)\,.
\end{align} 

As the local site group for the uranium atoms is equal to the isogonal point group, Table ML-I describes the local symmetry for the different IR. Then it is clear that only $A_{1}$ and $A_{2}$ are compatible with the fourfold rotational symmetry of the tetragonal space group. However, this symmetry is recovered also for the other IR if super-cells are allowed for.
This comes from the non-symmorphic group element $\left(c_{4}|\frac{1} {2}\frac{1} {2}\frac{1} {2}\right)$, a fourfold screw operation that connects the corner ($n$ even) and body-centered ($n$ odd) uranium sites in the original bct structure. It maintains the tetragonal symmetry also for the other IR, as for $B_{1}$ and $B_{2}$ we get
\begin{align} 
\left(c_{4}|\frac{1} {2}\frac{1} {2}\frac{1} {2}\right)\, {w}^{kpr}_{\pm 2}(0)&=-{w}^{kpr}_{\pm 2}(1)  \label{2-stagg}\\
\left(c_{4}|\frac{1} {2}\frac{1} {2}\frac{1} {2}\right)^{2}\, {w}^{kpr}_{\pm 2}(0)&={w}^{kpr}_{\pm 2}(2)
\end{align}
while for $E$ we get ($q$ is an integer $0\leq q \leq (r-1)/2$)
\begin{align} 
\left(c_{4}|\frac{1} {2}\frac{1} {2}\frac{1} {2}\right)\, {w}^{kpr}_{[q]}(0)&= {w}^{kpr}_{-[q]}(1)\\
\left(c_{4}|\frac{1} {2}\frac{1} {2}\frac{1} {2}\right)^{2}\, {w}^{kpr}_{[q]}(0)&= -{w}^{kpr}_{[q]}(2) \label{4-stagg}\\
\left(c_{4}|\frac{1} {2}\frac{1} {2}\frac{1} {2}\right)^{3}\, {w}^{kpr}_{[q]}(0)&=-{w}^{kpr}_{-[q]}(3) \\
\left(c_{4}|\frac{1} {2}\frac{1} {2}\frac{1} {2}\right)^{4}\, {w}^{kpr}_{[q]}(0)&= {w}^{kpr}_{[q]}(4)\,. 
\end{align}
where for odd $t=[q]=2q+1$, ${w}^{kpr}_t$ and ${w}^{kpr}_{-t}$ span the two-dimensional IR.
The operations leading to a change of sign of the TMTM, i.e.~Eqs.~(\ref{2-stagg}) and (\ref{4-stagg}), correspond to ordering vectors
$\vec{Q}=(001)$ and $\vec{Q}/2$, respectively.

\subsection{Susceptibility}
As discussed in the ML there are no experimental signature that the tetragonal crystal symmetry is broken in the hidden order (HO) phase.
However, recently in measurements of the in-plane susceptibility on single {\Ru} crystals an anisotropic component was detected. \cite{Okazaki}
This was subsequently analyzed by Thalmeier and Thakimoto	 \cite{Thalmeier}. Here we will extend  their analysis and show that the experiments can be explained by a non-vanishing 
$B_{2g}$ contribution to the in-plane susceptibility. 
The experiments are based on measurements of the torque on the sample in a uniform external magnetic field $H$ that are constrained to the $ab$ plane of the tetragonal crystal. This torque is given by
\begin{align} 
\vec{\tau}=\mu_\mathrm{B}V \chi \vec{H} \times \vec{H}\,,
\end{align}
where $V$ is the sample volume, $\chi$ the susceptibility tensor and $\vec{H}=\left(H_{a},H_{b},0\right)=H\left(\cos{\phi},\sin{\phi},0\right)$.
The in-plane part of the symmetric $\chi$ tensor is decomposed into three IR
\begin{align} 
{\mathcal Rep}\left[\chi_{\|}\right]=A_{1g}\oplus B_{1g}\oplus B_{2g}= 
{\mathcal Rep}\left[\chi_{aa}+\chi_{bb}\right]
\oplus {\mathcal Rep}\left[\chi_{aa}-\chi_{bb}\right]
\oplus{\mathcal Rep}\left[ \chi_{ab}\right]\,.
\end{align}
Then the torque is given by
\begin{align} 
\vec{\tau}=\mu_\mathrm{B}V H^{2} \hat{c} \left\{\frac{1}{2}(\chi_{aa}-\chi_{bb}) \sin{2\phi} - \chi_{ab}\cos{2\phi}    \right\}\,,
\end{align}
i.e.~with two-fold symmetric contributions from the $B_{1g}$ and $B_{2g}$ IR of the susceptibility, respectively.

Furthermore, these susceptibilities will have contributions from the HO, $\Psi(\vec{Q})$, through its interaction with the magnetic field in the Landau free energy expansion
\begin{align} 
{\mathcal F}_{H}&=g_{1}[\Psi(\vec{Q})^{2}]_{B_{1g}} [\vec{H}^{2}]_{B_{1g}}+g_{2}[\Psi(\vec{Q})^{2}]_{B_{2g}} [\vec{H}^{2}]_{B_{2g}}+ \dots\nonumber\\
 &= g_{1}[\Psi(\vec{Q})^{2}]_{B_{1g}} \frac{1}{2}(H_{a}^{2}-H_{b}^{2})+ g_{2}[\Psi(\vec{Q})^{2}]_{B_{2g}} H_{a}H_{b}+\dots\,,
\end{align}
where only second order interactions appear due to the staggering of the HO, $\vec{Q}\neq 0$, while the magnetic field is uniform.
Thus the $B_{1g}$ susceptibility are proportional to $[\Psi(\vec{Q})^{2}]_{B_{1g}}$, i.e.~the $B_{1g}$ part of the IR decomposition of the squared OP,  while the $B_{2g}$ susceptibility are proportional to $[\Psi(\vec{Q})^{2}]_{B_{2g}}$, since by definition $\chi_{ij}={\partial^{2}\mathcal F}/{\partial H_{i}\partial H_{j}}$.
From Table \ref{productIR} one can see that there are three ways each to get $B_{1g}$ and $B_{2g}$ contributions in the IR decomposition of a squared OP, i.e.~${\mathcal Rep}[\Psi(\vec{Q})]\otimes{\mathcal Rep}[\Psi(\vec{Q})]$.
$B_{1g}$ can be obtained from $A_{1\nu}\otimes B_{1\nu}$, $A_{2\nu}\otimes B_{2\nu}$ or $E_{\nu}\otimes E_{\nu}$, while 
$B_{2g}$ can be obtained from $A_{1\nu}\otimes B_{2\nu}$, $A_{2\nu}\otimes B_{1\nu}$ or $E_{\nu}\otimes E_{\nu}$, where $\nu$ indicates the TR symmetry, even ($g$) or odd ($u$). 

In the torque measurements \cite{Okazaki} a $\cos 2\phi$ oscillation was observed in the HO phase only, which imply that there is a non-vanishing $\chi_{ab}$, i.e.~a $B_{2g}$ IR, in the presence of the HO OP. Hence, if this OP belongs to a single IR of the point group $D_{4h}$, only  ${\mathcal Rep}[\Psi(\vec{Q})]=\Psi_{E_{\nu}}(\vec{Q})$ gives rise to the observed two-fold in-plane susceptibility $\chi_{ab}$, i.e.~only OP of IR $E_{g}$ or $E_{u}$ are candidates for the HO. On the other hand if the OP is a superposition of components belonging to different IR, OP of type  ${\mathcal Rep}[\Psi(\vec{Q})]=
\Psi_{A_{2\nu}}(\vec{Q})\oplus \Psi_{B_{1\nu}}(\vec{Q})\oplus\dots$ or ${\mathcal Rep}[\Psi(\vec{Q})]=
\Psi_{A_{1\nu}}(\vec{Q})\oplus \Psi_{B_{2\nu}}(\vec{Q})\oplus\dots$ are also possible.
Hence in a general case, with an OP having contributions from all possible IR, we have that
\begin{align} 
\chi_{ab}=\frac{\partial^{2}\mathcal F}{\partial H_{a}\partial H_{b}}=\sum_{\nu}2g_{2}\left(\Psi_{A_{1\nu}}\Psi_{B_{2\nu}}+\Psi_{A_{2\nu}}\Psi_{B_{1\nu}}+\Psi_{E_{\nu a}}\Psi_{E_{\nu b}}\right)\,.
\end{align}
Here $\Psi_{E_{\nu a}}$ and $\Psi_{E_{\nu b}}$ span the two-dimensional IR $E_{\nu}$ and are chosen such that $c_{4}E_{\nu a}=E_{\nu b}$, $c_{2}E_{\nu a}=(-)^{r+1}E_{\nu a}$ and $c_{2}E_{\nu b}=(-)^{r}E_{\nu a}$, with $r$ the rank of the OP tensor.

\begin{table}
\centering 
\caption{The IR decomposition of the direct product of two IR: the factors given by the row and column and the direct product in their intersection. 
$\nu$ labels the TR symmetry, either even $g$ or odd $u$.}
\begin{tabular} {|l||l|l|l|l|l|}
\hline  
$D_{4h}$ & $A_{1\nu}$& $A_{2\nu}$& $B_{1\nu}$ & $B_{2\nu}$ &  $E_{\nu}$ \\
\hline
$A_{1\nu}$ &  $A_{1g}$& $A_{2g}$& $B_{1g}$ & $B_{2g}$ &  $E_{g}$ \\
$A_{2\nu}$ &  $A_{2g}$& $A_{1g}$& $B_{2g}$ & $B_{1g}$ &  $E_{g}$ \\
$B_{1\nu}$ &  $B_{1g}$& $B_{2g}$& $A_{1g}$ & $A_{2g}$ &  $E_{g}$ \\
$B_{2\nu}$ &  $B_{2g}$& $B_{1g}$& $A_{2g}$ & $A_{1g}$ &  $E_{g}$   \\
$E_{\nu}$ & $E_{g}$& $E_{g}$& $E_{g}$ & $E_{g}$ &  $A_{1g}\oplus A_{2g}\oplus B_{1g}\oplus B_{2g}$ \\
\hline 
\end{tabular}
\label{productIR}
\end{table}

\subsection{Relativistic effects}

 In presence of strong spin-orbit coupling, there will be a large splitting between the $j=\ell-1/2=5/2$ and $j=\ell+1/2=7/2$ states.
Then it is useful to study the density matrix in a $j$$j$-basis
\begin{align} 
\rho_{n}=\left(\begin{array}{cc}
\rho_{n}^{\frac{5}{2}\frac{5}{2}} & \rho_{n}^{\frac{5}{2}\frac{7}{2}} \nonumber\\
\rho_{n}^{\frac{7}{2}\frac{5}{2}} & \rho_{n}^{\frac{7}{2}\frac{7}{2}} \end{array} \right) \,,\label{rhojj}
\end{align} 
where each sub-matrix $\rho_{n}^{j_{1}j_{2}}$ is spanned by $m_{1}$ and $m_{2}$ with $-j_{1}<m_{1}<j_{1}$ and $-j_{2}<m_{2}<j_{2}$.
\subsubsection{Strong spin-orbit coupling}
Let us first discuss the limit of very strong spin-orbit coupling. In this limit the TMTM $w^{110}_{0}$ is related to $w^{000}_{0}$ through  $w^{110}_{0}=-\frac{4}{3}w^{000}_{0}$, and only the $j=5/2$ occupation is non-zero. As we well discuss below, {\Ru} does not fulfill this criterion and hence rather possess an intermediately strong spin-orbit coupling.
However it is interesting to study this limit not least since it is  assumed in some other theoretical studies.
In this relativistic limit  the $j=7/2$ states are much higher in energy than the $j=5/2$ states
and the occupation is restricted to the sub-matrix $\rho_{n}^{\frac{5}{2}\frac{5}{2}}$ 
of Eq.~(\ref{rhojj}).
In this limit the TMTM are then given by
\begin{align} 
{w}^{kpr}_{t}(n)&\approx \mathrm{Tr} \, {\Gamma}^{kpr}_{t}\rho_{n}^{\frac{5}{2}\frac{5}{2}}\\
&=
\frac{6}{N_{kpr \ell}}
(-)^{k+p+r}
 \ninej{\ell} {\ell} {k} {s} {s} {p} {5/2} {5/2} {r} \, \sum_{m_{1}m_{2}} \gamma_{t,m_{1}m_{2}}^{\frac{5}{2}\frac{5}{2}r} \rho_{n,m_{2}m_{1}}^{\frac{5}{2}\frac{5}{2}}\,.\label{rellim}
\end{align}
Now from the definition of $\gamma^{jjr}_{t}$ in Eq.~(\ref{gamma}), one can see that $r$ in Eq.~(\ref{rellim}) is given by the vector coupling of two $j=5/2$ angular momenta,
and hence can take values in between 0 and 5. 
From general  relations for exchange of two columns in the $9j$ symbols 
\begin{align} 
 \ninej{\ell} {\ell} {k} {s} {s} {p} {j} {j} {r} = (-)^{2\ell+2s+2j+k+p+r}  \ninej{\ell} {\ell} {k} {s} {s} {p} {j} {j} {r}\,.\label{9jsymm}
\end{align}  
Now the fact that $2(\ell+s+j)$ is always even the $9j$-symbol in Eq.~(\ref{rellim}) has to be zero for odd $k+p+r$. Hence only even $k+p+r$ TMTM contribute in this limit.
This in turn  gives that 
odd (even) $r$ multipolar moments have to be TR-odd (TR-even). 

The different rank $r$ TMTM are then related through Eq.~(\ref{rellim}), e.g.~in the case of $r=5$ the TMTM with $kpr=505$, $415$ and $615$ have fixed ratios of
\begin{align}
\frac{w^{415}_{t}}{w^{615}_{t}}&=\frac{N_{615}}{N_{415}}\frac{ \ninej{3} {3} {4} {1/2} {1/2} {1} {5/2} {5/2} {5}}{ \ninej{3} {3} {6} {1/2} {1/2} {1} {5/2} {5/2} {5}}\approx0.013\label{eq415}\\
\frac{w^{505}_{t}}{w^{615}_{t}}&=\frac{N_{615}}{N_{505}}\frac{ \ninej{3} {3} {5} {1/2} {1/2} {0} {5/2} {5/2} {5}}{ \ninej{3} {3} {6} {1/2} {1/2} {1} {5/2} {5/2} {5}}\approx-0.077\,.\label{eq505}
\end{align}
In this case the $615$ TMTM is largest which is just an example of the general case. The spin dependent ($p=1$) TMTM have largest weight for the $r=k-1$ tensor moments, as these are favored by the spin-orbit coupling for less than half-filled $f$-shell.

\subsubsection{Intermediately strong spin-orbit coupling}
For a full $f$-shell the TMTM OP with highest rank is the $\psi^{617}_{t}(\vec{Q})$ 
of rank 7 and in this case the TR symmetry is not given by rank $r$, as also 
odd $k+p+r$ TMTM are allowed. 
They are
 included in all the analysis for completeness, although these tensor moments  are somewhat obscure. Within these the orbital and spin degrees of freedom couple in an axial way, 
as of e.g.~${\mathbf w}^{111}=\frac{2} {3}\vec{\ell}\times\vec{s}$. 
These tensor moments only arise from the off diagonal $j_{1}\neq j_{2}$ blocks of 
Eq.~(\ref{rhojj}) as they are not allowed in the block-diagonal part as the corresponding $9j$-symbols have to vanish according to Eq.~(\ref{9jsymm}).
They will be included in the analysis and it would be fascinating if they would play a role, but in the present case of {\Ru} they are of marginal interest.

That {\Ru} belongs to the case of intermediately strong spin-orbit coupling can be directly seen from the occupation numbers.
Both the $f$-occupation, $w^{000}_{0}$, as well as the effective spin orbit coupling, $w^{110}_{0}$, are essentially independent on the assumed IR
and take values around 2.6 and $-2.5$, respectively. This corresponds to that $w^{110}_{0}$ has a value $70\%$ of its saturation value, $-\frac{4}{3}w^{000}_{0}$.
The occupation number for the $5/2$ sub-shell is then given by
\begin{align} 
n_{5/2}=\frac{3}{7}w^{000}_{0}-\frac{3}{7}w^{110}_{0}\approx2.2\,,
\end{align}
while for the $7/2$ it is
\begin{align} 
n_{7/2}=\frac{4}{7}w^{000}_{0}+\frac{3}{7}w^{110}_{0}\approx0.4\,.
\end{align}

The extra flexibility of having a non-zero $n_{7/2}$ can lead to a stronger $615$-polarization on the expense of the $415$ and $505$ ones. This is e.g. confirmed in that the calculated ratios corresponding to Eqs.~(\ref{eq415}) and (\ref{eq505}) are $0.006$ and $-0.052$, respectively, for the case of $A_{1u}$ and $t=-4$.

\subsection{Polarization}
In order to quantify the importance of different tensor components we have utilized the concept of polarization~\cite{Polarisation}, $\pi^{kpr}_{t}$
 is a normalization independent quantity that directly measures the importance of the different contributions to the polarization of the density matrix. It is proportional to the square of the components of the TMTM $w^{kpr}_{t}$ and all components except $kpr=000$
 For each tensor component except, $k=p=r=0$,
\begin{align} 
\pi^{kpr}_{t}=[\ell skpr] |N_{kpr\ell}\,{w}^{kpr}_{t}|^2\,,
\end{align}
which all 
add up to a total polarization $\pi^\mathrm{tot}=\sum_{kprt}\pi^{kpr}_{t}$. The total polarization is constrained by the inequality $\pi^\mathrm{tot}\leq nn_{h}$, where $n$ is the occupation number of the $f$-shell and $n_{h}$ the corresponding number of holes.

\section{Electronic structure calculations}
In ML a systematic study which of the TMTM are compatible with the electronic structure were conducted through realistic calculations.

The electronic structure was determined with the DFT+$U$ approximation within the APW+$lo$ method as implemented in the {\sc elk}-code. \cite{LAPW,Elk}
{In the DFT+$U$ approach a screened Hartree-Fock interaction is included among the 5$f$ states only. The Slater parameters $F^{(k)}$ are determined individually by using a screened Yukawa potential, this approach has been showed to be particularly convenient since all four $F^{(k)}$ are controlled by the choice of a single parameter, 
the parameter $U$. \cite{Bultmark-Mult}
For double-counting we adopt the automatic interpolation scheme of Pethukov {\em et al}. \cite{Pethukov-INT}
The calculations were performed at the experimental lattice constants. \cite{Palstra} 
The muffin-tin radius $R^{U}_{\mathrm{MT}}$ of U is set to 1.9~\AA, and those of
Ru and Si are set to 1.2 {\AA}. The parameter $R^{U}_{\mathrm{MT}} |\vec{G}+\vec{k}|_{\mathrm{max}}$, governing  the number of plane waves in the APW+$lo$ method, is chosen to be 9.5. To allow for the $\vec{Q}$ order the unit cell is doubled and the corresponding Brillouin zone is sampled with $18\times18\times10$ $k$-points.}

Care has to be taken to allow for TMTM OP solutions as of Eq.~(\ref{genOP}) that belong to the different IR of the isogonal group $D_{4h}$ as listed in Table I of ML. In our approach we break the symmetry by inducing staggered  $w^{kpr}_{t}$ on the uranium sites and determine the largest possible symmetry group that is compatible with their existence. Then by iteration we determine if this starting assumption converges to a non-trivial solution.

\end{document}